\documentclass[epj]{svjour}

\usepackage{graphicx}
\usepackage{amsmath}
\usepackage{ulem}
\usepackage{color}
\newcommand{\PGRcomm}[1]{{\color{black} #1}}
\newcommand{\MDcomm}[1]{{\color{black} #1}}

\begin{document}
\title{Exploration of dynamical regimes of irradiated small protonated
  water clusters}
\author{
Uguette Flore\ Ndongmouo Taffoti\inst{1}
\and Phuong Mai\ Dinh \inst{2}
\and Paul-Gerhard\ Reinhard\inst{2,3}
\and Eric\ Suraud\inst{2}
\and Zhi Ping\ Wang\inst{4}
}
\institute{
CEA Saclay, DEN/SRMP, F-91191 Gif/Yvette Cedex
\and 
Laboratoire de Physique Th\'{e}orique, IRSAMC, UPS and CNRS,
Universit\'{e} de Toulouse, 118 Rte de Narbonne, F-31062 Toulouse
cedex, France
\and
Institut f{\"{u}}r Theoretische Physik, Universit{\"{a}}t Erlangen,
Staudtstrasse 7, D-91058 Erlangen, Germany 
\and School of Science, JiangNan University, Wuxi 214122, China
}
\date{First draft: oct 2009}
%
\abstract{
  We explore from a theoretical perspective the dynamical response of small
  water clusters, (H$_2$O)$_n$H$_3$O$^+$ with $n=1,2,3$, to a short laser
  pulse for various frequencies, from infrared (IR) to ultra-violet (UV) and
  intensities (from $6\times10^{13}$ W/cm$^2$ to $5\times10^{14}$
  W/cm$^2$). To that end, we use time-dependent local-density
  approximation for the electrons, 
  coupled to molecular dynamics for the atomic cores (TDLDA-MD).  The
  local-density approximation is augmented by a self-interaction correction
  (SIC) to allow for a correct description of electron emission.
  For IR frequencies, we see a direct coupling of the laser field to
  the very light H$^+$ ions in the clusters. Resonant coupling (in the UV)
  and/or higher intensities lead to fast ionization with subsequent
  Coulomb explosion. The stability against Coulomb pressure increases with
  system size. Excitation to lower ionization stages induced
  strong ionic vibrations. These maintain rather harmonic pattern
  in spite of the sizeable amplitudes (often 10\% of the bond length).
%
} 
\maketitle

\section{Introduction}

The structure of water is generally considered to be rather ordered
due to hydrogen bonds and with certain structures prevailing.  There
is, in turn, a general agreement among models about the existence of
small and medium-sized clusters of water molecules \cite{Nov99}.
Furthermore, experimental evidence point to the fact that under
appropriate conditions, water clusters may ionize.  There thus exists,
since rather long, a large body of investigations on charged water
clusters, either anions \cite{New75,Hab84,Bar88}, cations \cite{Tom83}
or protonated clusters \cite{New77,Yan89}.

The dynamics of water clusters appears as a key issue in various scenarios,
e.g., for the chemistry of formation of droplets and clouds in atmosphere
\cite{Ste01}.  Water clusters are also possible candidates as transient
intermediate stages in liquid water \cite{Sti80}. They might thus play a key
role in many chemical and physical processes
\cite{Bjo99,Mac00,Xan00,Rap03,Jor04aR}.  The case of charged
clusters, and especially ionized clusters, is particularly relevant for
studies on radiation damage, as water is the natural "environment" of
biomolecules. This has triggered recent studies on the influence of a finite
number of water molecules coating a biomolecule \cite{Liu06}. The direct
analysis of the irradiation of protonated water clusters is also a key
issue~\cite{Ado09}. 

From the theoretical point of view, water clusters have been considered since
long at a structural level \cite{New75,Bar88,Ven07}, for a recent summary see
\cite{Zwi04aR}. In recent years, studies of irradiation of biomolecules,
possibly in a water environment have also been started \cite{Gai07,Koh08}, but
usually without an explicit time-dependent treatment of the ionization process
itself.  In the present paper, we aim at studying the response of small
protonated water clusters to an irradiation by a short laser
pulse. The choice of 
laser irradiation rather than excitation by a charged projectile is to some
extent a matter of convenience as it allows for exploratory studies a
discrimination and a systematic scan of various dynamical regimes without
dealing with complications of scattering geometry. Furthermore, there exists a
well documented literature on the laser irradiation of clusters, metallic or
rare gas mostly \cite{Saa06,Fen10} in a wide range of excitations, which
provides a robust background to study irradiation of other sorts of clusters
such as water clusters. The paper is organized as follows. We first briefly
recall the basic inputs of the model, in section \ref{sec:model}.  As a next
step we first consider the optical response which is known to provide a key
entry point to understand irradiation dynamics (section \ref{sec:opt}). We
then address irradiation by an intense laser filed, exploring in particular
the influence of laser frequency and intensity as key characteristics of the
laser pulse (section \ref{sec:las}).  We finally draw our conclusions in
section \ref{sec:conc}.

\section{Model}
\label{sec:model}

In order to perform microscopic simulations of dynamical processes, we
employ time-dependent density functional theory
(TDDFT)~\cite{Run84,Gro90,Mar04} 
for the valence electrons (described in terms of single-electron
wave functions $\{\varphi_\alpha({\bf r}), \alpha=1 \ldots N_{\rm
el}\}$) combined with classical molecular dynamics (MD) for the ionic
cores (described by their classical coordinates $\{{\mathbf R}_I,I=1
\ldots N_{\rm ion}\}$).  The ions are here H$^+$, and O$^{6+}$.
The starting point is an energy which includes the electron kinetic
energy, the direct Coulomb energy, the exchange-correlation energy,
the coupling energy between electrons and ions (described in terms of
non-local pseudo-potentials of the type \cite{Goe96a} and tuned to
provide a most efficient description on the numerical
grid~\cite{Wan10}), and the energy
coming from the interaction of the system with the external
time-dependent laser field coupling to both, electrons and ions. The
functional for the electronic exchange-correlation energy is taken at
the level of the local-density approximation (LDA) with the actual
form from~\cite{Per92}. The electronic energy is augmented with an
average-density self-interaction correction (ADSIC) \cite{Leg02}.
The ADSIC is a crucial ingredient to provide correct ionization
potentials (IP) which, in turn, is necessary to describe electron
emission correctly.

Equations of motion are then derived in a standard manner by variation
of the given energy. Variation with respect to $\varphi_\alpha$
leads to the (time-dependent) Kohn-Sham equations for the
single-particle wave functions.  Variation with respect to the ionic
coordinates yields Hamiltonian equations of motion for ions, thus
treated by classical MD.
Electronic and ionic equations of motion are solved simultaneously
leading to coupled TDLDA-MD description \cite{Cal00,Rei03a}. 
The treatment goes thus beyond the adiabatic approximation.
It becomes equivalent to the Born-Oppenheimer (adiabatic)
approximation in regimes where the latter is valid. 

Time-dependent fields and wave functions are represented on a 3D
coordinate-space grid of dimensions $72 \times 72 \times 64$ or
$72^3$, with a grid spacing of 0.412 $a_0$. The electronic wave
functions are 
propagated in time by the time-splitting method~\cite{Fei82}. The
Coulomb problem is solved by a fast Fourier
technique~\cite{Lau94}. The ionic equations of motion are solved using
the Verlet algorithm. Absorbing boundary conditions are used to
properly remove outgoing electrons \cite{Cal00,Rei03a}.  Thus the
total number of electrons $N=N(t)$ decreases in time. The number of
escaped electrons $N_\mathrm{esc}=N(t\!=\!0)-N(t)$ is a measure of
average ionization.

\begin{figure}[htbp]
\begin{center}
\includegraphics[width=\linewidth]{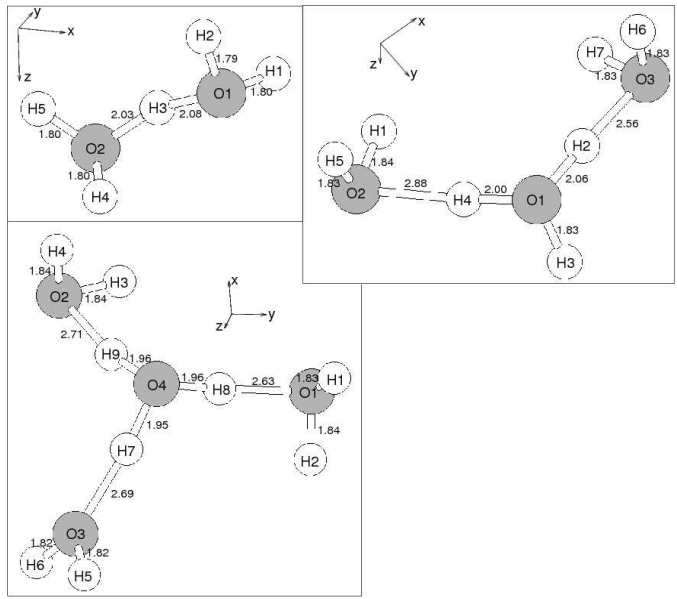}
\caption{Schematic view of the initial configuration of the systems
  H$_2$O\,H$_3$O$^+$, (H$_2$O)$_2$\,H$_3$O$^+$, and
  (H$_2$O)$_3$\,H$_3$O$^+$. Note that all indicated frames are
  positively oriented.} 
\label{fig:geom}
\end{center}
\end{figure}
The electronic ground state wave functions are obtained by damped
gradient iteration \cite{Blu92}. Ionic ground state configurations are
obtained by cooling pseudo-dy\-na\-mics \cite{Car85}.  Results for the
structure of a few selected water clusters are shown in figure
\ref{fig:geom}. Gray balls stand for the oxygen atoms and are labelled
O1, O2, etc. White ball, labelled H1, etc., represent the hydrogen
atoms. The O-H bond lengths are also indicated and take basically three
values, around 1.8, 2.0 and 2.7 $a_0$.

\section{Optical response}
\label{sec:opt}

The spectral distribution of optical absorption strength (optical
response) provides key information on the coupling of a system to
(laser) light.  The initial response is mostly mediated by
electrons. The optical response then is related to eigenfrequencies
of electronic excitations which exhibit well isolated peaks at
frequencies below emission threshold and continuum above.  These
eigenmodes are important because they represent the doorways for
laser excitation.  In the case of metal clusters, the major absorption
strength lies in the optical domain with a marked collective
character. They are known as the Mie plasmon resonances and correspond
to a collective oscillation of the electron cloud with respect to the
ionic background, with rather well established systematic behaviors
with cluster size \cite{Rei03a,Bra93}.  The situation is more involved
in organic systems where varying bonding can lead to large differences
from one cluster to the next. Furthermore the typical eigenfrequencies
lie in the UV part of spectrum rather than in the optical domain and
thus require laser pulses with higher frequencies (single photon) or
higher intensities (multiphoton processes).

The optical absorption strength can be computed easily within our
TDLDA description by the technique of spectral analysis
\cite{Cal95a,Yab96}, for a detailed discussion see \cite{Cal97b}. To
that end, we apply initially an infinitely short pulse covering all
frequencies with equal weight, propagate the system for a sufficiently
long time (depending on the desired spectral resolution), record the
time-dependent dipole momentum $\mathbf D(t)$, and finally Fourier transform
the signal into the frequency domain.  The strength is then obtained
as the imaginary part of $\tilde{\mathbf{D}}(\omega)$.
%
\begin{figure}[htbp]
\begin{center}
\includegraphics[width=\linewidth]{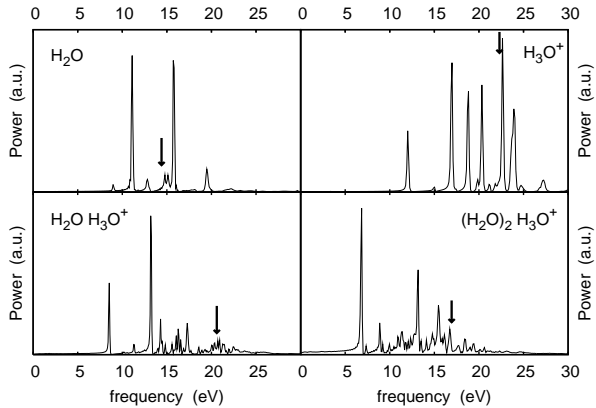}
\caption{Spectral distribution of optical
absorption  strength for the systems H$_2$O, H$_3$O$^+$,
H$_2$O\,H$_3$O$^+$, and (H$_2$O)$_2$\,H$_3$O$^+$.
The corresponding IP are indicated by an arrow.
} 
\label{fig:opt}
\end{center}
\end{figure}
Results for optical absorption spectra of H$_2$O, H$_3$O$^+$,
H$_2$O\,H$_3$O$^+$, and (H$_2$O)$_2$\,H$_3$O$^+$ are shown in Figure
\ref{fig:opt}. 
The spectra exhibit typical structures with well marked and isolated
eigenfrequencies at "low" frequencies below the ionization potential (IP) for
the given system and a trend to a more fuzzy "continuum" response at higher
frequencies above the IP.  The figure shows that infrared and visible light is
below any significant peak and thus off resonant. Light pulses in the UV range
can couple more or less strongly, dependening an the detailed frequency.
The figure does also indicate the IP for each system. These are in detail:
IP(H2O) $=-14.4$ eV, IP(H$_3$O$^+$) $=-22.3$ eV,
IP(H$_2$O\,H$_3$O$^+$) $=-20.6$ eV, and 
IP((H$_2$O)$_2$H$_3$O$^+$) $=-16.9$ eV.

\section{Irradiation by a strong laser field}
\label{sec:las}

We now turn to direct analysis of the irradiation and response of small
protonated water clusters to short and intense laser pulses.  Even in
the UV regime, the laser wavelength remains much larger than the size
of the system under consideration. So the long wavelength limit
applies such that the dipole operator describes the spatial part
of laser pulse.  Thus we model the laser pulse as 
the time dependent external field
\begin{equation}
  V_{\rm las} 
  =
  E_0  \mathbf{e}_\mathrm{pol}\cdot\hat{\mathbf{r}} \cos(\omega t)f(t)
\label{eq:laser}
\end{equation}
with the amplitude \PGRcomm{$E_0$ being related to the intensity $I$
as} \MDcomm{$E_0(\mbox{Ry}.a_0^{-1}) = 1.07 \times 10^{-8} \sqrt{I
(\mbox{W/cm}^2)}$} and (linear) polarization
$\mathbf{e}_\mathrm{pol}$.  \MDcomm{In all our calculations, we use
$\mathbf{e}_\mathrm{pol}= \mathbf{e}_x$ (see Figure
\ref{fig:geom}). We will see below, \PGRcomm{in the discussion of
  figure \ref{fig:bond},} that the laser polarization has
only a limited importance \PGRcomm{for the basic mechanisms.}}
%
The laser frequency $\omega$ is carried in the cos
term.  The pulse profile $f(t)$ is chosen as a sin$^2$ characterized
by its Full Width at Half Maximum (FWHM). In all calculations below we
have chosen a FWHM of 20 fs and we have focused on the simple case of
(H$_2$O)H$_3$O$^+$, exploring only at some places larger clusters.

\subsection{Dependence on frequency}
\begin{figure}[htbp]
\begin{center}
\includegraphics[width=\linewidth]{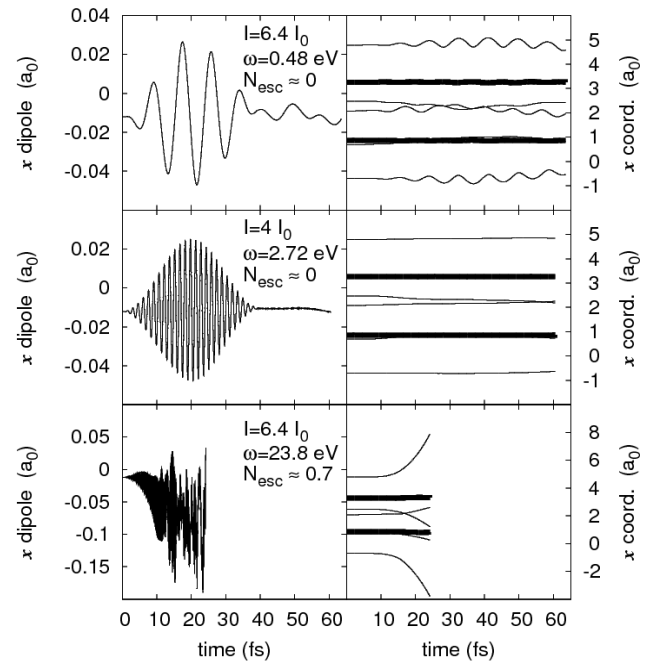}
\caption{Irradiation of (H$_2$O)H$_3$O$^+$ by lasers with polarization
along $x$ axis, FWHM of 20 fs, intensities of about $I_0=10^{13}$
W/cm$^2$, and three frequencies as indicated.  The number of escaped
electrons $N_{\rm esc}$ is also reported for each case. As a function
of time~: Electronic dipole 
moment in $x$ direction (left column), and ionic $x$ coordinates
(right column). The thick curves stand for the position of the oxygen
atoms, while the thin lines show the $x$ coordinates of the hygrogen
atoms.}  
\label{fig:w}
\end{center}
\end{figure}

We first consider the influence of laser frequency on irradiation
dynamics of (H$_2$O)H$_3$O$^+$, exploring a variation of laser
frequency at basically constant laser intensity.  Figure \ref{fig:w}
shows results for three representative cases: a frequency in the
infra-red (IR), another one in the visible range, and an ultra-violet
(UV) one. The IR and visible frequencies lie below the range of
typical electronic eigenfrequencies in (H$_2$O)H$_3$O$^+$ (see Figure
\ref{fig:opt}), while the UV case lies in the middle of the
eigenfrequencies. The overall dynamical behavior is strongly related
to the laser frequency. We observe a typical out-of-resonance behavior
for IR and visible frequencies where the amplitude of the dipole
signal (left column of Figure \ref{fig:w}) is proportional to the
envelope of the laser field. There is practically no electron
emission for the given laser intensity.  The UV frequency, being in
resonance with some of the systems excitation modes, displays a much
different behavior: the strong dipole response lasts longer than the
laser signal with a sizeable ionization. Correspondingly, the
overall behavior of the cluster dynamics behaves quite
differently. The right column of Figure \ref{fig:w} displays the
time evolution of the ionic positions of constituting ions along the
laser polarization axis. While the two off-resonance cases exhibit
only low amplitude ionic vibrations, the UV case leads to an almost
immediate explosion of the (H$_2$O)H$_3$O$^+$.

Another interesting aspect concerns the difference between the visible
and IR cases. In the frequency in the visible range, one can hardly
see any ionic perturbation while the IR irradiation leads to clear
ionic vibrations with an small, but not vanishing, amplitude.  This
difference can be traced back to a direct coupling of the ions
(predominantly the hydrogen ions) with the laser itself. This seems a
bit surprising in view of the rather small FWHM of the laser pulse (20
fs). But the small hydrogen mass allow such a rapid coupling. This
coupling is observed in the IR domain and not for the visible light
because typical ionic vibration frequencies in small water clusters
precisely lie in the considered IR frequency domain \cite{Yu07}.
Estimating the H vibration frequency in this case leads to a value of
$\hbar \omega \sim 0.41$~eV, in fair agreement with previously published
results which shows peaks around 0.47~eV \cite{Yu07}.
Closer inspection shows that the IR laser pulse
simply excites internal vibrations of hydrogen atoms mostly along
OH bonds.  

In order to explore this effect in more detail, we now focus on the distances
between O and H atoms. Indeed, because of their larger masses, O atoms
basically remain fixed during the short time scales explored here and we can
thus focus the further analysis on O--H bonds only.  
\begin{figure}[htbp]
\begin{center}
\includegraphics[width=\linewidth]{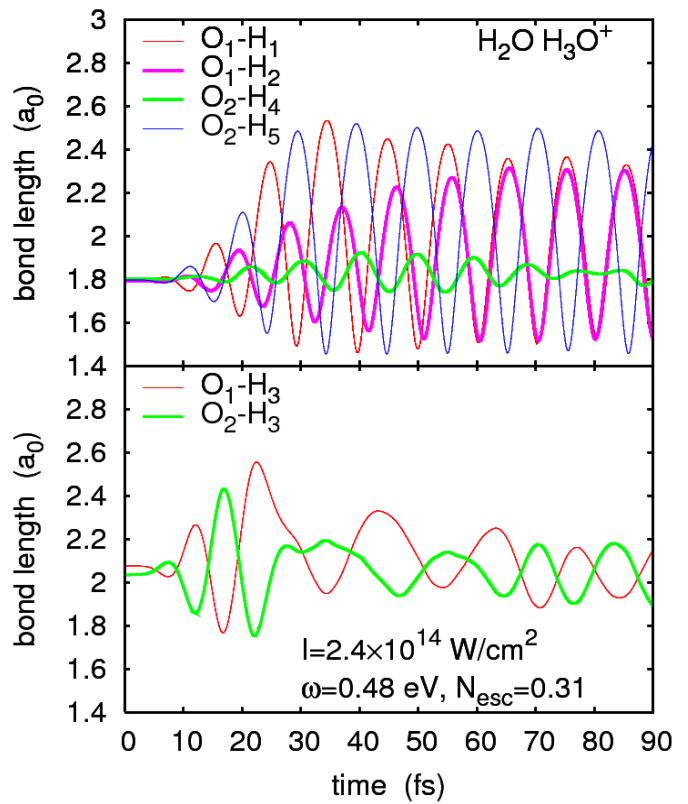}
\caption{Irradiation of (H$_2$O)H$_3$O$^+$ by a laser with
polarization along $x$ axis, FWHM of 20 fs, and intensity and
frequency as indicated. The number of escaped electrons is also
given. Are plotted as a function of time the O--H bond lengths~: the 
shortest ones (top), according to the atom numbering of
Figure~\ref{fig:geom}, and the longest ones (bottom).}
\label{fig:bond}
\end{center}
\end{figure}
As can be seen from
Figure \ref{fig:geom}, there are basically two types of O--H bonds, the ones
between O and external H and the internal ones along the sequence H--O--H.
The associated bond lengths are different and so are the corresponding
vibration frequencies. It is thus interesting to consider these two classes of
cases separately. The results are plotted in Figure
\ref{fig:bond}. One should first note the remarkable regularity of
oscillations, especially for external H atoms (upper panel), with about all
having the same frequency. One also clearly observes oscillations in opposite
phases depending on the geometry.  This is particularly obvious for the two
internal bonds (lower panel of Figure \ref{fig:bond}).  Different heights of
amplitudes indicates a different strength of the response which is
predominantly a geometry effect due to the orientation of the bond relative to
laser polarization, \MDcomm{here the $x$ direction. Indeed, the
O$_1$-H$_1$ and O$_2$-H$_5$ bonds are the external O-H bonds which are
aligned with the $x$ axis the most (see Figure~\ref{fig:geom}) and
which exhibit the largest amplitudes of oscillations. Changing the
laser polarization (along the $y$ or the $z$ direction for instance)
would produce a largest excitation of other O-H bonds but the
mechanisms of excitation would remain the same.}
Finally one should note the amplitude of oscillations 
which may be rather large, typically 10\% of the actual bond length but
without destroying the harmonicity of the oscillations, the latter aspect
being especially true for external H atoms.

\subsection{Dependence on intensity}

We have identified in Figure \ref{fig:w} the resonant coupling of the
laser pulse with internal IR vibrations along OH bonds and explored it 
in more detail in Figure \ref{fig:bond}. 
\begin{figure}[htbp]
\begin{center}
\includegraphics[width=\linewidth]{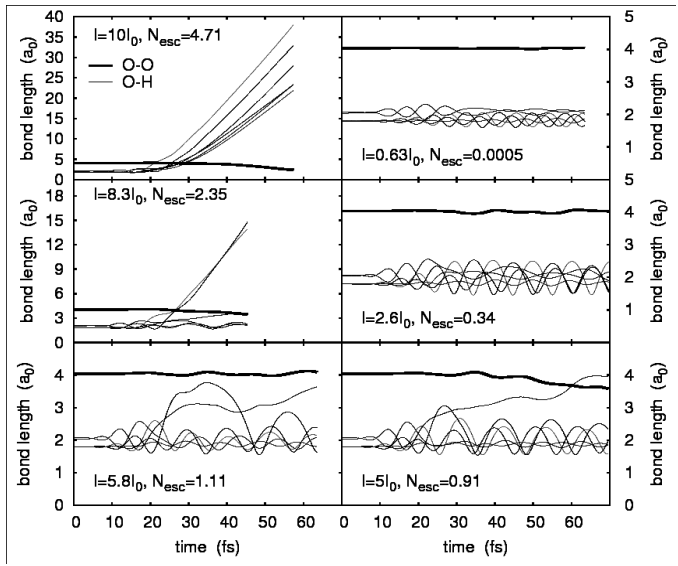}
\caption{Irradiation of (H$_2$O)H$_3$O$^+$ by lasers with polarization
along $x$ axis, FWHM of 20 fs, frequency of 0.48 eV, and 
six different intensities
varying about $I_0=10^{14}$ W/cm$^2$ as indicated.
Intensities increase in clockwise order.
 The average
ionization (number of escaped
electrons $N_{\rm esc}$) is also indicated for each case. Various bond
lengths (O-O and shortest O-H) are depicted as a function
of time.}
\label{fig:I}
\end{center}
\end{figure}
Figure \ref{fig:I} explores how this behavior depends on laser
intensity. It 
shows the time evolution of the distances between the 2 oxygen atoms
which lies around 4 $a_0$ and the typical OH bonds whose length is
about 2 $a_0$. The influence of laser intensity on the ionic response
is obvious. With increasing laser intensity, we observe increasing
ionization. Up to about net ionization one, there is little effect on
the general vibrational pattern. They gain, of course, amplitude due
to the larger field strength. But on the time scales computed here, the
systems seems to resist fragmentation and keep its vibrational
response. Only for ionization substantially larger than one, the dynamical
evolution changes qualitatively to the extent that the now large
repulsion leads to an immediate fragmentation of the cluster and, of
course, disappearance of the vibrational response. All in all, we thus
observe a remarkable robustness of the vibrational response with
increasing laser intensity.

\subsection{Dependence on size}

\begin{figure}[htbp]
\begin{center}
\includegraphics[width=\linewidth]{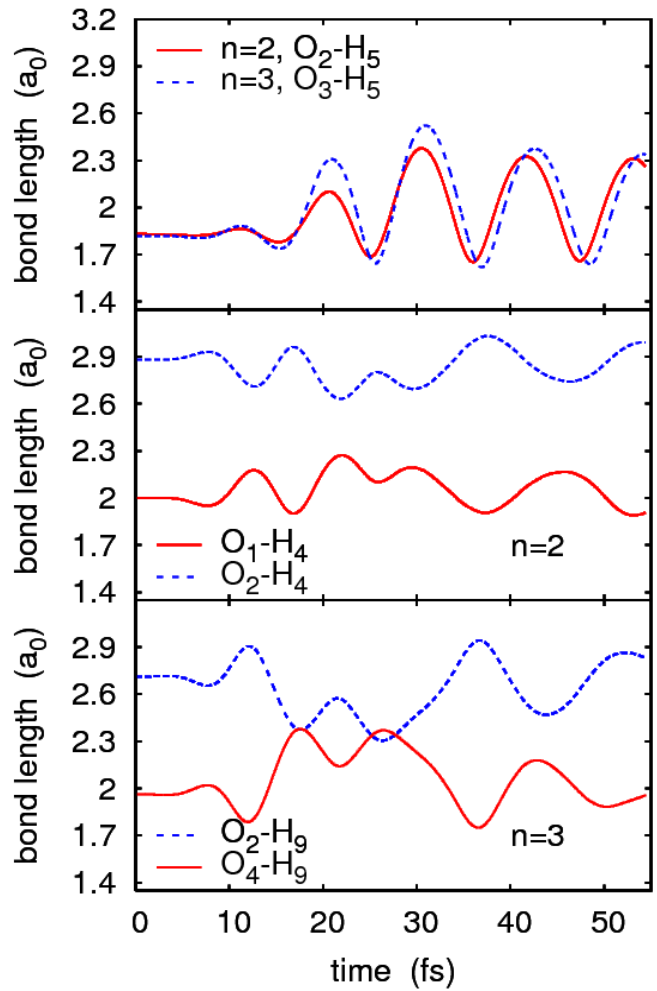}
\caption{Irradiation of (H$_2$O)$_n$H$_3$O$^+$, for $n=2,3$ by
lasers with polarization along $x$ axis, FWHM of 20 fs, frequency
of 0.48 eV, and intensity of $2.6 \times 10^{14}$ W/cm$^2$. Various
O--H bond lengths are depicted as a function of time, with labels
according to the atom numbering of Figure~\ref{fig:geom}.} 
\label{fig:size}
\end{center}
\end{figure}

It is finally interesting to explore how dynamical evolution depends
on cluster size. We consider here the case of (H$_2$O)$_n$H$_3$O$^+$
clusters for $n=1,2,3$.  
We again take the same IR frequency at 0.48
eV and a moderate laser intensity to remain in the vibrational regime
in (H$_2$O)H$_3$O$^+$. Various OH bond lengths are plotted as a
function of time in Figure~\ref{fig:size}. Because of the quickly 
increasing number of OH bonds with increasing size, which makes
 a figure including all of them 
unreadable, we have selected a few representative cases, following the 
labelling given in Figure \ref{fig:geom}. Note that, although 
we use the same laser in all cases, the net ionization differs from one case 
to the next, due to different ionization potentials:  
$-20.6$~eV, $-16.9$eV, and $-15.7$~eV for $n$=1, 2, and 3
respectively; associated net ionizations are 0.34, 1.44, 2.00
respectively for $n=1,2,3$. 
This should not be a problem as we have seen that the effects are 
to a large extent robust with net ionization. 
Indeed, at least on the short time scale we
have considered here, we observe in all cases comparable vibrational
response, in spite of the large ionization in the largest
cluster. Note, however, that the larger ionization appears in the
larger system which has naturally a higher capability to cope with
a high charge state. Such an effect has been found also for
the Coulomb fragmentation of metal clusters \cite{Nae97}. 

Figure~\ref{fig:size} exhibits interesting features in detail. 
The upper panel compares typical  external O--H vibrations 
in the $n=2$ and $n=3$. In both clusters, the corresponding 
bonds have very similar lengths and frequencies, 
as is clear form the plot. The responses are remarkably similar. 
The middle panel focuses on (H$_2$O)$_2$H$_3$O$^+$ internal
H--O--H sequence comparing H--O and O--H vibrations, which 
turn out to be pretty similar and nicely in phase opposition, 
simply because the O atoms basically do not move. Finally 
the lower panel concerns (H$_2$O)$_3$H$_3$O$^+$, again 
focusing on a H--O--H sequence and with the same conclusions 
as in the middle panel. 

\section{Conclusion}
\label{sec:conc}

We have explored in this paper the dynamical response of small water
clusters (in particular (H$_2$O)H$_3$O$^+$) to a short and intense
laser pulse for various frequencies and intensities covering the regime
of stable vibrations up to Coulomb break-up of a highly ionized
cluster. The first stages of response of the system are, as is well
known from other systems, primarily taken up by the electrons and we
have recovered this behavior.  For low frequencies in the infra-red
(IR), we also see a direct coupling of the laser pulse to the ionic
centers of the clusters, particularly to the most mobile H$^+$ ions.
This coupling becomes sizeable because the ionic vibrations are close
to resonant conditions for IR frequencies.  The effect is robust with
respect to change of laser intensity, even for sizeable extra
ionization up to typically one more charge unit.  
We have also studied larger hydrogen clusters with one or two more
water molecules and found similar behaviors. We see here that the
limits of stability against Coulomb break-up is shifted to higher
ionization with increasing system size.
The statements about stability have to be taken with care because our
observation times are rather short. This means that we have discussed
at least a stable transient stage. One needs to follow the evolution
for longer time spans to find out more about the long-time stability.
Work along that line is in progress.

\begin{acknowledgement}
This work was supported by the 
{Deutsche Forschungsgemeinschaft (RE 322/10, RO 293/27),}
Fonds der Chemischen Industrie (Germany), a Bessel-Humboldt prize,
a Gay-Lussac prize, the French Agence Nationale de la Recherche
(ANR-06-BLAN-0319-02), and the French computational facilities CalMip
(Calcul en Midi-Pyr\'en\'ees), IDRIS, CINES and CCRT.
\end{acknowledgement}

\bibliographystyle{epj}

\begin{thebibliography}{99}

\bibitem{Nov99}
Yulia V. Novakovskaya, Nikolai F. Stepanov, J. Phys. Chem. A {\bf 103}
(1999) 3285.


\bibitem{New75}
M. D. Newton, J. Phys. Chem. {\bf 79} (1975) 2795.

\bibitem{Hab84}
H. Haberland, C. Ludewigt, H.-G. Schindler, D. R. Worsnop, J.
Chem. Phys. {\bf 81} (1984) 3742.

\bibitem{Bar88}
R. N. Barnett, U.  Landman, C. L. Cleveland, J. Jortner, J. Chem.
Phys. {\bf 88} (1988) 4429.

\bibitem{Ven07}
O. Vendrell, F. Gatti, and H.-D. Meyer, Angew. Chem., Int. Ed. {\bf 46} (2007) 6918.


\bibitem{Tom83}
S. Tomoda, K. Kimura, Chem. Phys. Lett. {\bf 102} (1983) 560.

\bibitem{New77}
M. D. Newton, J. Chem. Phys. {\bf 67} (1977) 5535.


\bibitem{Yan89}
X. Yang, A. W. Jr. Castleman, J. Am. Chem. Soc. {\bf 111} (1989) 6845.

\bibitem{Ste01}
Z. Sternovsky, M. Hor´anyi, S. Robertson, Phys. Rev. A {\bf 64} (2001)
023203. 

\bibitem{Sti80}
F. H. Stillinger, Science {\bf 209} (1980) 451.

\bibitem{Bjo99} 
O. Bj\"orneholm, F. Federmann, S. Kakar, and T. M\"oller
J. Chem. Phys. {\bf 111} (1999) 546.

\bibitem{Mac00} 
R.S. MacTaylor, A.W. Castleman Jr., J. Atmos. Chem. {\bf 36} (2000)
23.

\bibitem{Xan00}
S.S. Xantheas (Ed.), Recent Theoretical and Experimental
Advances in Hydrogen-bonded Clusters, NATO Science Series
C, vol. 561, Kluwer, Dordrecht, 2000.

\bibitem{Rap03}
M. Rapp, F.-J. Lu¨bken, T.A. Blix, Adv. Space Res. {\bf 31} (2003)
2033.

\bibitem{Jor04aR}
K. D. Jordan,
Science {\bf 306} (2004) 618.

\bibitem{Liu06}
B. Liu, S. Br\o ndsted Nielsen, P. Hvelplund, H. Zettergren,
H. Cederquist, B. Manil, and B. A. Huber, Phys. Rev. Lett. {\bf 97}
(2006) 133401. 

\bibitem{Ado09}
L. Adoui, A. Cassimi, B. Gervais, J.-P. Grandin, L. Guillaume, R.
Maisonny, S. Legendre, M. Tarisien, P. L\'opez-Tarifa, M.-F. Politis, M.-A.
Herv\'e du Penhoat, R. Vuilleumier, M.-P. Gaigeot, I. Tavernelli,
M. Alcam\'\i, and F. Mart\'\i n, 
J. Phys. B {\bf 42} (2009) 075101.

\bibitem{Zwi04aR}
T. S. Zwier, 
Science {\bf 304} (2004) 1119.

\bibitem{Gai07}
M. P. Gaigeot, R. Vuilleumier, C. Stia, M. E. Galassi, R. Rivarola,
B. Gervais, M. F. Politis, 
J. Phys. B  {\bf 40} (2007) 1.

\bibitem{Koh08}
J. Kohanoff, E. Artacho, AIP Conf Proc {\bf 1080} (2008) 78.

\bibitem{Saa06}
U. Saalmann, C. Siedschlag, J. M. Rost,
{J. Phys. B} {\bf 39} (2006) R39.

\bibitem{Fen10}
Th. Fennel, K.--H. Meiwes-Broer, J. Tiggesb\"aumker, P.~M. Dinh,
P.--G. Reinhard, E. Suraud, 
Rev. Mod. Phys. (2009) in press.

\bibitem{Run84}
E. Runge, E. K. U. Gross, Phys. Rev. Lett. {\bf 52} (1984) 997.

\bibitem{Gro90}
E. K. U. Gross, W. Kohn,
Adv. Quant. Chem. {\bf 21} (1990) 255.

\bibitem{Mar04}
M. A. L. Marques, E. K. U. Gross,
Ann. Rev. Phys. Chem. {\bf 55} (2004) 427.

\bibitem{Goe96a}
S. Goedecker, M. Teter, and J. Hutter,
Phys. Rev. B {\bf 54} (1996) 1703.

\bibitem{Wan10}
Z. P. Wang, P. M. Dinh, P.-G. Reinhard, E. Suraud,
in preparation.

\bibitem{Per92}
J. P. Perdew and Y. Wang,
Phys. Rev. B {\bf 45} (1992) 13244.

\bibitem{Leg02}
C. Legrand, E. Suraud, P.G. Reinhard, J. Phys. B {\bf 35}, (2002) 1115.

\bibitem{Cal00}
F.~Calvayrac, P.-G. Reinhard, E. Suraud, C.A. Ullrich, Phys. Rep.
{\bf 337} (2000) 493.

\bibitem{Rei03a}
P.-G. Reinhard, E. Suraud, \emph{Introduction to Cluster Dynamics}
(Wiley, New York, 2003).



\bibitem{Fei82}
M. D. Feit, J. A. Fleck, and A. Steiger,
J. Comp. Phys. {\bf 47} (1982) 412.

\bibitem{Lau94}
G. Lauritsch and P.-G. Reinhard,
Int. J. Mod. Phys. C {\bf 5} (1994) 65.

\bibitem{Blu92}
V. Blum, G. Lauritsch, J. A. Maruhn, and P.-G. Reinhard,
{J. Comp. Phys} {\bf 100} (1992) 364.

\bibitem{Car85}
R. Car and M. Parinello,
Phys. Rev. Lett. {\bf 55} (1985) 2471.

\bibitem{Bra93}
M.~Brack,
Rev. Mod. Phys. {\bf 65} (1993) 677.

\bibitem{Cal95a}
F. Calvayrac, E. Suraud, P.-G. Reinhard,
Phys. Rev. B {\bf 52} (1995) R17056

\bibitem{Yab96}
K.~Yabana and G.~F. Bertsch,
Phys. Rev. B {\bf 54} (1996) 4484.

\bibitem{Cal97b}
F. Calvayrac, P.-G. Reinhard, E. Suraud,
Ann. Phys. (NY) {\bf 255} (1997) 125.


\bibitem{Yu07}
Haibo Yu and Qiang Cui,
J. Chem. Phys. {\bf 127} (2007) 234504

\bibitem{Nae97}
U. N\"aher, S. Bj\"ornholm, S. Frauendorf, F. Garcias, and C. Guet,
Phys. Rep. {\bf 285} (1997) 245.


\end{thebibliography}

\end{document}